\documentclass[12pt,a4paper]{article}
\usepackage{amsmath,epsfig}
\renewcommand{\theequation}{\arabic{section}.\arabic{equation}}

\begin{document}
 
\setlength{\unitlength}{.8mm}
\begin{titlepage} 
\vspace*{0.2cm}
\begin{center}
{\Large\bf TBA equations for the mass gap in the O$(2r)$ 
non-linear $\sigma$-models}
\end{center}
\vspace{1.5cm}
\begin{center}
{\large J\'anos Balog$^{1,3}$ and \'Arp\'ad Heged\H us$^{2,3}$}
\end{center}
\bigskip

{\it 
\begin{center}
$^1$ Institut f\"ur Physik, Humboldt-Universit\"at zu Berlin, \\ 
Newtonstr. 15, 12489 Berlin, Germany         
\end{center}

\medskip
\begin{center}
$^2$ Istituto Nazionale di Fisica Nucleare, Sezione di Bologna, \\
Via Irnerio 46, 40126 Bologna, Italy
\end{center}

\medskip
\begin{center}
$^3$ Research Institute for Particle and Nuclear Physics,\\
Hungarian Academy of Sciences,\\
H-1525 Budapest 114, P.O.B. 49, Hungary\\ 
\end{center}
}
\vspace{2.2cm}
\begin{abstract}
We propose TBA integral equations for 1-particle 
states in the O$(n)$ non-linear $\sigma$-model for even $n$. 
The equations are conjectured on the basis of the analytic properties
of the large volume asymptotics of the problem, which is explicitly
constructed starting from L\"uscher's asymptotic formula.
For small volumes the mass gap values computed numerically from the 
TBA equations agree very well with results of three-loop perturbation
theory calculations, providing support for the validity of the
proposed TBA system.
\end{abstract}

\end{titlepage}

\section{Introduction and Summary}
The two-dimensional integrable O$(n)$ non-linear $\sigma$-models have been the 
subject of a huge amount of study because they often arise in 
experimentally-realizable condensed-matter systems, because their 
integrability offers powerful theoretical methods applicable and because 
they share a lot of common properties with QCD, such as renormalizability, 
asymptotic freedom, dynamical mass generation, and instanton solutions.
A particular interest of these models appears in lattice field theory, where 
because of their low dimensionality, the O$(n)$ non-linear $\sigma$-models are 
used to test new ideas.
 
Many integrable quantum field theories (including the O$(n)$ non-linear 
$\sigma$-models) have been exactly solved \lq\lq on-shell" i.e. their mass 
spectra and their factorizable S-matrices have been determined \cite{Zamo}. 
Important progress was made towards the \lq\lq off-shell" solution of the 
integrable quantum field theories by the so called Thermodynamic Bethe Ansatz 
(TBA) technique \cite{3BLZ} which proved to be very successful~\cite{4BLZ}.
This approach allows one to calculate the ground state energy of an 
integrable quantum field theory enclosed in a finite box with periodic 
boundary conditions, provided that the \lq\lq on-shell" solution is known. 
As the result of this method one obtains a set of coupled nonlinear integral 
equations (TBA equations), by the solutions of which the ground state 
energy can be determined. The knowledge of the finite size dependence of the 
ground state energy makes it possible to probe the scale dependence in 
integrable quantum field theories.
 
The derivation of the TBA equations is cumbersome in the case of the O$(n)$ 
non-linear $\sigma$-models because of the difficulties appearing in the 
classification of the solutions of the Bethe Ansatz equations in the 
thermodynamical limit. This problem has been solved by formulating 
the O$(n)$ non-linear $\sigma$-models as limits of certain perturbed conformal 
field theories \cite{f0,f1,f2,f3,Fen}, where this classification can be done 
making possible the derivation of TBA equations for these models.
The $\sigma$-model TBA equations are the limits of those of the corresponding 
perturbed conformal field theories.
 
At the same time it seems to be interesting to calculate the full finite size
spectrum of the O$(n)$ non-linear $\sigma$-models, because these energies 
would provide interpolation between the small volume (perturbative) spectrum 
and the large volume (massive) spectrum of the theory. 
 
To be able to achieve this, one needs to generalize the TBA technique
for excited states. Although a lot of different methods have been worked out to
obtain excited state TBA equations in different models 
\cite{exTBA0,exTBA1,exTBA2,exTBA3,exTBA4,exTBA5}, a general method has not been
discovered yet. The derivation of functional equations for the eigenvalues of 
certain mutually commuting transfer matrices is one of the most powerful 
method to attack the excited state problem in integrable quantum field 
theories \cite{exTBA2,exTBA5}. Knowing the analytic properties of these 
eigenvalues the functional equations can be transformed into integral 
equations \cite{89EB} which generalize the standard ground state TBA equations 
to excited states. This has been worked out for some simple perturbed 
conformal field theories directly in the continuum \cite{exTBA2} or starting 
from the integrable lattice regularization of the model \cite{exTBA5}.
   
Unfortunately in the case of the O$(n)$ non-linear $\sigma$-models it is not 
obvious how to derive such functional equations (if they exist), and how to 
get information about the analytic properties of the functions in the case of 
excited states. 
   
As a partial solution of the problem, in \cite{BH2} a simple method was worked 
out for proposing one-particle excited state TBA equations in the special 
cases of the O$(3)$ and O$(4)$ non-linear $\sigma$-models. The purpose of our 
work here is to extend this method to the case of general O$(n)$.

In this paper we propose an (infinite) set of TBA integral equations
describing the 1-particle excitation energies in the O$(n)$ non-linear
$\sigma$-model for even $n$, defined in a finite periodic 1-dimensional
box. The equations are not derived from first principles and their validity
is based on two major assumptions. First, we assume that the excited states
are associated with the same Y-system equations as the ones proposed earlier
for the ground state of the system \cite{Fen}. Second, we assume that
the analytic structure of the Y-system functions is independent (at least
qualitatively) of the system size and can be read off from the infinite
volume solution, which can be constructed using L\"uscher's asymptotic 
formula \cite{Luscher}. The first assumption can be proved for the sine-Gordon
model \cite{BH1} and (at least for a subset of the physical states of the 
system) also for other simple models including the O$(4)$ $\sigma$-model.
The second assumption is difficult to prove, even for the sine-Gordon
model. However, in our previous studies \cite{BH1,BH2} we verified the
validity of these assumptions by showing that the TBA equations obtained
using them give finite volume energy values that agree very well with
the results of Monte Carlo simulations and (for small volumes) with
the results of perturbative calculations. In the same spirit, we study 
the finite volume mass gap values of the $O(n)$ models (for $n=6$ and $8$)
numerically here and show that again, for small volumes, the agreement 
with the results of perturbation theory is excellent. This can be taken 
as an indirect proof of our assumptions. Indeed, since our starting point
is the large volume asymptotic formula, the fact that the mass gap is
correctly given by our equations also for small volumes is a very strong 
indication for the overall correctness of our approach.

The paper is organized as follows. In Section~2 we recall the TBA integral
equations describing the finite volume physics of the ground 
state of the model and write down the corresponding TBA Y-system equations.
In Section~3, using L\"uscher's formula and assuming that the same Y-system
describes the excited states, we formulate the 1-particle TBA problem.
We find the solution in the infinite volume limit explicitly for 
$O(4)$ and O$(6)$ and conjecture the explicit form of the infinite volume  
solution for general rank. In Section~4 we prove this conjecture.
In Section~5, assuming that the analytic structure of the solution is
always the same as in the infinite volume limit, we transform the 1-particle
Y-system into TBA integral equations. The numerical solution of these 
equations is briefly discussed in Section~6 and the mass gap values are 
compared to perturbative results. The discussion of some technical details
are relegated into the Appendices. Appendix~A discusses a TBA lemma that
enables one to translate TBA integral equations into TBA Y-systems and
vice versa. Some useful formulae related to the O$(n)$ S-matrix are collected
in Appendix~B. In Appendix~C the solution of the constant Y-system is
given together with that of the related Q-system. Finally in Appendix~D
the details of the proof of Section~4 can be found. 

\setcounter{equation}{0}
\section{O$(2r)$ TBA system}
\newsavebox{\ON}
\sbox{\ON}{\begin{picture}(52,5)(0,-3.5)

\put(30,20){\circle*{3}}
\put(30,18.5){\line(0,-1){17}}
\put(29,18.7){\line(-1,-1){18.0}}
\put(29.5,18.6){\line(-1,-2){8.7}}
\put(31,18.7){\line(1,-1){18.0}}
\put(30.5,18.6){\line(1,-2){8.7}}
\put(30.7,19.6){\line(3,-1){27.7}}
\put(30.7,19.1){\line(5,-2){36.1}}

\put(10,0){\circle{3}}
\multiput(20,0)(10,0){4}{\circle{3}}
\multiput(11.5,0)(10,0){2}{\line(1,0){7}}
\multiput(32.5,0)(1,0){6}{\circle*{.2}}
\put(41.5,0){\line(1,0){7}}
\put(51.1,1.1){\line(1,1){7.7}}
\put(51.4,0){\line(4,1){15.2}}
\put(60,10){\circle{3}}
\put(68,4){\circle{3}}

\put(10,-1.5){\line(0,-1){7}}
\put(20,-1.5){\line(0,-1){7}}
\put(30,-1.5){\line(0,-1){7}}
\put(40,-1.5){\line(0,-1){7}}
\put(50,-1.5){\line(0,-1){7}}
\put(60,8.5){\line(0,-1){7}}
\put(68,2.5){\line(0,-1){7}}

\put(10,-11.5){\line(0,-1){7}}
\put(20,-11.5){\line(0,-1){7}}
\put(30,-11.5){\line(0,-1){7}}
\put(40,-11.5){\line(0,-1){7}}
\put(50,-11.5){\line(0,-1){7}}
\put(60,-1.5){\line(0,-1){7}}
\put(68,-7.5){\line(0,-1){7}}

\put(10,-21.5){\line(0,-1){7}}
\put(20,-21.5){\line(0,-1){7}}
\put(30,-21.5){\line(0,-1){7}}
\put(40,-21.5){\line(0,-1){7}}
\put(50,-21.5){\line(0,-1){7}}
\put(60,-11.5){\line(0,-1){7}}
\put(68,-17.5){\line(0,-1){7}}

\put(10,-31.5){\line(0,-1){2}}
\put(20,-31.5){\line(0,-1){2}}
\put(30,-31.5){\line(0,-1){2}}
\put(40,-31.5){\line(0,-1){2}}
\put(50,-31.5){\line(0,-1){2}}
\put(60,-21.5){\line(0,-1){2}}
\put(68,-27.5){\line(0,-1){2}}

\multiput(10,-34.8)(0,-1){3}{\circle*{.2}}
\multiput(20,-34.8)(0,-1){3}{\circle*{.2}}
\multiput(30,-34.8)(0,-1){3}{\circle*{.2}}
\multiput(40,-34.8)(0,-1){3}{\circle*{.2}}
\multiput(50,-34.8)(0,-1){3}{\circle*{.2}}
\multiput(60,-24.8)(0,-1){3}{\circle*{.2}}
\multiput(68,-30.8)(0,-1){3}{\circle*{.2}}

\put(10,-10){\circle{3}}
\multiput(20,-10)(10,0){4}{\circle{3}}
\multiput(11.5,-10)(10,0){2}{\line(1,0){7}}
\multiput(32.5,-10)(1,0){6}{\circle*{.2}}
\put(41.5,-10){\line(1,0){7}}
\put(51.1,-8.9){\line(1,1){7.7}}
\put(51.4,-10){\line(4,1){15.2}}
\put(60,0){\circle{3}}
\put(68,-6){\circle{3}}

\put(10,-20){\circle{3}}
\multiput(20,-20)(10,0){4}{\circle{3}}
\multiput(11.5,-20)(10,0){2}{\line(1,0){7}}
\multiput(32.5,-20)(1,0){6}{\circle*{.2}}
\put(41.5,-20){\line(1,0){7}}
\put(51.1,-18.9){\line(1,1){7.7}}
\put(51.4,-20){\line(4,1){15.2}}
\put(60,-10){\circle{3}}
\put(68,-16){\circle{3}}

\put(10,-30){\circle{3}}
\multiput(20,-30)(10,0){4}{\circle{3}}
\multiput(11.5,-30)(10,0){2}{\line(1,0){7}}
\multiput(32.5,-30)(1,0){6}{\circle*{.2}}
\put(41.5,-30){\line(1,0){7}}
\put(51.1,-28.9){\line(1,1){7.7}}
\put(51.4,-30){\line(4,1){15.2}}
\put(60,-20){\circle{3}}
\put(68,-26){\circle{3}}


\end{picture}}

In this section we summarize the TBA integral equations proposed by
Fendley \cite{Fen} for the ground state energy of the O$(n)$ non-linear
$\sigma$-models for even $n$, $n=2r$. It is formulated in terms
of a finite TBA system describing a certain perturbed conformal field
theory model that becomes the O$(2r)$ $\sigma$-model in the limit of
infinitely many TBA components.

The finite problem is characterized by the real pseudo-densities
$\epsilon_0(\theta)$ and $\epsilon^{(a)}_m(\theta)$. Here the upper index
$a=1,\dots,r$ labels the nodes of the $D_r$ Dynkin-diagram ($r\geq2$) and 
the lower index runs in the range $m=1,\dots,k-1$, where $k\geq2$ is the 
Kac-Moody level characterizing the conformal limit of the finite system.
The $\sigma$-model corresponds to the limit $k\to\infty$ \cite{Fen}.

We now make the following definitions.
\begin{equation}
Y_0(\theta)={\rm e}^{-\epsilon_0(\theta)},\qquad
L_0=\ln[1+Y_0(\theta)]
\end{equation}
and
\begin{equation}
\left.
\begin{aligned}
Y^{(a)}_m(\theta)={\rm e}^{-\epsilon^{(a)}_m(\theta)},\qquad
L^{(a)}_m(\theta)&=\ln[1+Y^{(a)}_m(\theta)],\\
 \Lambda^{(a)}_m(\theta)&=\ln\left[1+\frac{1}{Y^{(a)}_m(\theta)}\right],
\end{aligned}
\right\}
\begin{aligned}
a&=1,\dots,r,\\m&=1,\dots,k-1.
\end{aligned}
\end{equation}
There is a symmetry relation
\begin{equation}
\epsilon^{(r-1)}_m(\theta)=\epsilon^{(r)}_m(\theta),\qquad
m=1,\dots,k-1,
\end{equation}
which follows from the reflection symmetry of the $D_r$ Dynkin-diagram.
The ground state energy is given by
\begin{equation}
E_0(R)=-\frac{M}{2\pi}\,\int_{-\infty}^\infty\,{\rm d}\theta\,
\cosh\theta\,L_0(\theta),
\label{E0}
\end{equation}
where $R$ is the size of the (one-dimensional) periodic box and 
$M$ is the particle mass in infinite volume. We will also use the notation
$\ell=MR$ for the dimensionless size of the system.
The pseudo-densities have to be determined from the TBA integral equations
\begin{equation}
\epsilon_0(\theta)=MR\cosh\theta-\sum_{a=1}^{r-1}\,\int_{-\infty}^\infty\,
{\rm d}u\,{\cal R}\left(\frac{\pi}{2},\frac{(r-1-a)\pi}{g};\theta-u\right)\,
L_1^{(a)}(u)
\label{epsilon0}
\end{equation}
and for $a=1,\dots,r$, $m=1,\dots,k-1$
\begin{equation}
\begin{split}
-\epsilon_m^{(a)}(\theta)=\frac{1}{2}\,\int_{-\infty}^\infty\,
{\rm d}u\,&{\cal R}\left(\frac{\pi}{g},0;\theta-u\right)\,\Big\{
\delta_{m1}(\delta_{a1}+\delta_{a2}\delta_{r2})L_0(u)\\
&\qquad\quad+L^{(a)}_{m+1}(u)+L^{(a)}_{m-1}(u)
-\sum_{b=1}^r\,I_{ab}\Lambda^{(b)}_m(u)\Big\},
\end{split}
\label{epsilonam}
\end{equation}
where the kernel function ${\cal R}(\alpha,\beta;\theta)$ is given 
by (\ref{kernel}), $g=2(r-1)$ is the Coxeter number of the $D_r$ Lie-algebra
and (by definition)
\begin{equation}
L^{(a)}_0(\theta)=L^{(a)}_k(\theta)=0,\qquad a=1,\dots,r.
\end{equation}
Finally $I_{ab}$ is the incidence matrix of the $D_r$ Dynkin-diagram, which
is unity if the nodes $a,b$ are connected and is zero otherwise.

We can now use the TBA lemma discussed in Appendix A to transform the integral
equations into TBA Y-systems. We first do this for the equations (\ref{epsilonam}),
for which it is sufficient to use the special case of the lemma. We get
for $a=1,\dots,r$ and $m=1,\dots,k-1$
\begin{equation}
\begin{split}
&Y^{(a)}_m\left(\theta+\frac{i\pi}{g}\right)
Y^{(a)}_m\left(\theta-\frac{i\pi}{g}\right)=
\big\{1+\delta_{m1}(\delta_{a1}+\delta_{a2}\delta_{r2})Y_0(\theta)\big\}\\
&\qquad\quad\cdot
\big\{1+Y^{(a)}_{m+1}(\theta)\big\}\big\{1+Y^{(a)}_{m-1}(\theta)\big\}\,
\prod_{b=1}^r\left(\frac{Y^{(b)}_m(\theta)}{1+Y^{(b)}_m(\theta)}\right)^{I_{ab}}
\label{Y1}
\end{split}
\end{equation}
with boundary condition
\begin{equation}
Y^{(a)}_0(\theta)=Y^{(a)}_k(\theta)=0,\qquad a=1,\dots,r.
\end{equation}
For (\ref{epsilon0}) we have to use the lemma in its general form. We assume
that 
\begin{equation}
1+Y_1^{(a)}(\theta): {\rm \ ANZ}\qquad \vert{\rm Im}\,\theta\vert\leq
\frac{(r-1-a)\pi}{g},\qquad a=1,\dots,r-2,
\end{equation}
where ANZ indicates that the function is analytic and non-vanishing in the
given strip. If this is satisfied then we can use the TBA lemma to transform
(\ref{epsilon0}) into 
\begin{equation}
Y_0\big(\theta+\frac{i\pi}{2}\big)Y_0\big(\theta-\frac{i\pi}{2}\big)=
\prod_{a=1}^{r-1}
\Big\{1+Y^{(a)}_1\big(\theta+\frac{i(r-1-a)\pi}{g}\big)\Big\}
\Big\{1+Y^{(a)}_1\big(\theta-\frac{i(r-1-a)\pi}{g}\big)\Big\}.
\label{Y2}
\end{equation}
The TBA Y-system (\ref{Y1}), (\ref{Y2}) is illustrated in Figure 1. 
It belongs to the class of TBA Y-systems with one massive node 
($Y_0$ in this case).

\begin{figure}[htbp]
\vspace{0.6cm}
\begin{center}
\begin{picture}(140,40)(-18,-20)
\put(15,0) {\usebox{\ON}}
\put(-30,-45){\parbox{132mm}{\caption{\label{DpAsp}\protect {\footnotesize
Infinite Y-system corresponding to the O$(2r)$
non-linear $\sigma$-model }}}}
\end{picture}
\end{center}
\vspace{2.0cm}
\end{figure}
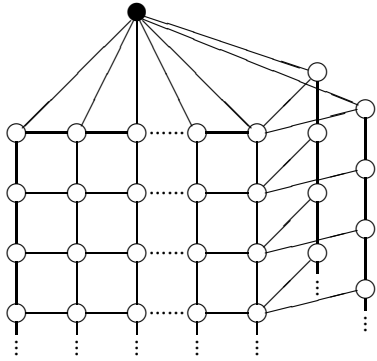

\setcounter{equation}{0}
\section{1-particle Y-system}
In this section we conjecture the structure of the 1-particle Y-system, 
generalizing previous examples \cite{BH1,BH2}, in particular the case 
of the O$(4)$ non-linear $\sigma$-model (corresponding to $r=2$). 
Our starting point is L\"uscher's asymptotic formula \cite{Luscher}. 
For large $\ell$
\begin{equation}
E_H(R)\approx HM-\frac{M}{2\pi}\int_{-\infty}^\infty\,{\rm d}\theta\,
\cosh\theta\,{\rm e}^{-\ell\cosh\theta}y(\theta)\qquad (H=0,1).
\label{Luscher}
\end{equation}
Here $H=0$ and $H=1$ correspond to the ground state energy and first excited 
state energy of the O$(n)$ $\sigma$-model, respectively and for
\begin{eqnarray}
H&=&0:\quad y(\theta)=n,\\
H&=&1:\quad y(\theta)=q\big(\theta+\frac{i\pi}{2}\big), \quad
{\rm where}\quad q(\theta)=\sum_{b=1}^n\,S^{ab}_{ab}(\theta).
\label{1party}
\end{eqnarray}
For $H=0$ (\ref{Luscher}) is giving the first (universal) term of the
virial expansion \cite{virial}, which is proportional to the number of 
particles ($n$), while for $H=1$ the coefficient function $y(\theta)$
can be obtained from the exact S-matrix of the model \cite{Zamo}.

On the other hand for a TBA system with a single massive node $Y_\bullet$ 
the analogous expressions are
\begin{equation}
E_H(R)\approx HM-\frac{M}{2\pi}\int_{-\infty}^\infty\,{\rm d}\theta\,
\cosh\theta\,\ln\{1+Y_\bullet(\theta)\}\qquad (H=0,1).
\label{EH}
\end{equation}
Comparing this with L\"uscher's formula we conclude that asymptotically
\begin{equation}
Y_\bullet(\theta)\approx{\rm e}^{-\ell\cosh\theta}y(\theta).
\end{equation}

In our case we conjecture that the 1-particle energy of the system is given by
\begin{equation}
E_1(R)= M-\frac{M}{2\pi}\int_{-\infty}^\infty\,{\rm d}\theta\,
\cosh\theta\,\ln\{1+Y_0(\theta)\}, 
\label{E1}
\end{equation}
where the 1-particle $Y_0$ satisfies the asymptotic condition
\begin{equation}
Y_0(\theta)\approx{\rm e}^{-\ell\cosh\theta}y(\theta)
\end{equation}
with the 1-particle $y(\theta)$ defined in (\ref{1party}) and together
with the other Y-functions $Y^{(a)}_m$ satisfy the ($k\to\infty$ limit
of the) Y-system equations (\ref{Y1}) and (\ref{Y2}). Actually the
structure of the equations becomes more transparent after introducing
the variables $X_0$, $X^{(a)}_m$ with rescaled arguments:
\begin{equation}
Y_0(\theta)=X_0\left(\frac{g\theta}{\pi}\right),\qquad\qquad
Y_m^{(a)}(\theta)=X_m^{(a)}\left(\frac{g\theta}{\pi}\right).
\end{equation}
In terms of these (\ref{Y1}) is rewritten as
\begin{equation}
\begin{split}
&X^{(a)}_m(x+i) X^{(a)}_m(x-i)=
\big\{1+\delta_{m1}(\delta_{a1}+\delta_{a2}\delta_{r2})X_0(x)\big\}\\
&\qquad\quad\cdot\big\{1+X^{(a)}_{m+1}(x)\big\}\big\{1+X^{(a)}_{m-1}(x)\big\}\,
\prod_{b=1}^r\left(\frac{X^{(b)}_m(x)}{1+X^{(b)}_m(x)}\right)^{I_{ab}}
\label{X1}
\end{split}
\end{equation}
for $a=1,\dots,r$, $m=1,2,\dots$ and (\ref{Y2}) becomes
\begin{equation}
X_0\big(x+i(r-1)\big)X_0\big(x-i(r-1)\big)=
\prod_{a=1}^{r-1}
\Big\{1+X^{(a)}_1\big(x+i(r-1-a)\big)\Big\}
\Big\{1+X^{(a)}_1\big(x-i(r-1-a)\big)\Big\}.
\label{X2}
\end{equation}

In order to be able to transform (\ref{X1}) and (\ref{X2}) into integral 
equations we need to know their analytic structure (position of poles, zeroes)
in the strip relevant for the application of the TBA lemma of Appendix A.
Our strategy now is to study the large $\ell$ limit of the system first and
try to establish the analytic structure of the Y-functions in this limit.
We assume that in this limit 
\begin{eqnarray}
X_0(x)&\approx&{\rm e}^{-\ell\cosh\frac{\pi x}{g}}y\big(\frac{\pi x}{g}\big),\\
X^{(a)}_m(x)&\approx&\bar X^{(a)}_m(x),
\end{eqnarray}
where $\bar X^{(a)}_m(x)$ is an $\ell$-independent function. It follows that
\begin{equation}
X_0\big(x+i(r-1)\big)X_0\big(x-i(r-1)\big)\quad\to\quad
y\big(\frac{\pi x}{g}+\frac{i\pi}{2}\big)
y\big(\frac{\pi x}{g}-\frac{i\pi}{2}\big)
\end{equation}
and using the results in Appendix B that
\begin{equation}
\begin{split}
\prod_{a=1}^{r-1}
&\Big\{1+\bar X^{(a)}_1\big(x+i(r-1-a)\big)\Big\}
\Big\{1+\bar X^{(a)}_1\big(x-i(r-1-a)\big)\Big\}\\
&\qquad\quad
=\frac{4r^2\gamma\big(x+i(r-1)\big)\gamma\big(x-i(r-1)\big)}
{(x^2+4)(x^2+g^2)},
\end{split}
\label{luscherX}
\end{equation}
where
\begin{equation}
\gamma(x)=x^2+\frac{(r-2)(r^2-1)}{r}.
\end{equation}
The $\ell\to\infty$ limit of (\ref{X1}) simplifies to ($a=1,\dots,r$
\ \  $m=1,2,\dots$)
\begin{equation}
\bar X^{(a)}_m(x+i) \bar X^{(a)}_m(x-i)=
\big\{1+\bar X^{(a)}_{m+1}(x)\big\}\big\{1+\bar X^{(a)}_{m-1}(x)\big\}\,
\prod_{b=1}^r\left(\frac{\bar X^{(b)}_m(x)}{1+\bar X^{(b)}_m(x)}\right)^{I_{ab}}
\label{barX1},
\end{equation}
which is the $k\to\infty$ limit of the standard $A_{k-1}\,\diamond\,D_r$
Y-system \cite{diamond}.

For the simplest case, $r=2$, we have to determine the functions
$\bar X^{(1)}_m$ for $m=0,1,\dots$. In this case (\ref{luscherX}) determines  
\begin{equation}
\bar X^{(1)}_1(x)=\frac{3x^2}{x^2+4}
\end{equation}
and using (\ref{barX1}) recursively leads to the complete solution
\begin{equation}
\bar X^{(1)}_m(x)=\frac{m(m+2)x^2}{x^2+(m+1)^2},\qquad\qquad
1+\bar X^{(1)}_m(x)=\frac{(m+1)^2(x^2+1)}{x^2+(m+1)^2}.
\end{equation}
We note that this is of the form 
\begin{equation}
\bar X^{(1)}_m(x)=y_m\frac{x^2}{x^2+1+y_m},\qquad\qquad
1+\bar X^{(1)}_m(x)=(1+y_m)\frac{x^2+1}{x^2+1+y_m},
\end{equation}
where $y_m$ are the constant solution of the $r=2$ Y-system (see Appendix C).

In the next example we are going to study, the $r=3$ case, there are two
sets of functions, $\bar X^{(1)}_m$ and $\bar X^{(2)}_m$,  $m=0,1,\dots$.
The general solution of (\ref{luscherX}) is
\begin{eqnarray}
1+\bar X^{(1)}_1(x)&=&\frac{9}{4}\,\frac{\gamma(x+i)\gamma(x-i)}{x^2(x^2+9)}\,
\frac{1}{\psi^2(x)}\\
1+\bar X^{(2)}_1(x)&=&\frac{8}{3}\,\frac{x^2+1}{\gamma(x)}\,
\psi(x+i)\psi(x-i),
\end{eqnarray}
where $\psi(x)$ is an arbitrary real analytic function.
For the simplest choice, $\psi(x)=1$, this can be extended to a solution
for all $m=0,1,\dots$
\begin{eqnarray}
\bar X^{(1)}_m(x)&=&\frac{m(m+4)}{4}\,
\frac{\big(x^2+b_{m+1}^2\big)\big(x^2+b_{m-1}^2\big)}
{x^2\big(x^2+(m+2)^2\big)},\label{D31}\\
\bar X^{(2)}_m(x)&=&\frac{m(m+4)}{3}\,
\frac{x^2}{x^2+b_m^2},\\
1+\bar X^{(1)}_m(x)&=&\frac{(m+2)^2}{4}\,
\frac{\big(x^2+(b_m+1)^2\big)\big(x^2+(b_m-1)^2\big)}
{x^2\big(x^2+(m+2)^2\big)},\label{D33}\\
1+\bar X^{(2)}_m(x)&=&\frac{(m+1)(m+3)}{3}\,
\frac{x^2+1}{x^2+b_m^2},
\end{eqnarray}
where
\begin{equation}
b_m^2=\frac{(m+1)(m+3)}{3}.
\end{equation}
Again, we can recognize the structure
\begin{equation}
\bar X^{(2)}_m(x)=y_m^{(2)}\frac{x^2}{x^2+1+y_m^{(2)}},\qquad\qquad
1+\bar X^{(2)}_m(x)=(1+y_m^{(2)})\frac{x^2+1}{x^2+1+y_m^{(2)}},
\end{equation}
where $y_m^{(2)}$ are the constant solution of the $D_3$ Y-system.

Based on these suggestive examples, we now formulate our conjecture for
general rank. For the infinite volume solution for general $r$ the functions 
associated to the \lq\lq fork" part of the Dynkin-diagram are given by 
\begin{equation}
\bar X^{(r-1)}_m(x)=\bar X^{(r)}_m(x)=y_m^{(r-1)}\frac{x^2}{x^2+1+y_m^{(r-1)}},
\label{con1}
\end{equation}
\begin{equation}
1+\bar X^{(r-1)}_m(x)=1+\bar X^{(r)}_m(x)
=(1+y_m^{(r-1)})\frac{x^2+1}{x^2+1+y_m^{(r-1)}},
\label{con2}
\end{equation}
where $y^{(r-1)}_m$ are the constant solution of the $D_r$ Y-system 
(see Appendix C). 
Starting from here, the Y-system equations (\ref{barX1}) determine
the remaining functions $\bar X^{(a)}_m(x)$ for $a=1,\dots,r-2$ and
$m=1,2,\dots$. Thus this solution, if exists, is unique. This will be
discussed in the next section.

\setcounter{equation}{0}
\section{Infinite volume solution}
In this section we give the complete solution of the infinite volume
Y-system explicitly extending the conjectures (\ref{con1}-\ref{con2}).
We start by introducing 
\begin{alignat}{3}
y^{(a)}_m\ &: \qquad & a&=1,\dots,r & \qquad m&=0,1,\dots,\\
p^{(a)}_m\ &: \qquad & a&=0,1,\dots,r-1 & \qquad m&=0,1,\dots,
\end{alignat}
where $y^{(a)}_m$ are the solution of the constant Y-system described
in Appendix C and  $p^{(a)}_m$ are some additional (real) parameters.
We also introduce the squares
\begin{equation}
\xi^{(a)}_m=\big(p^{(a)}_m\big)^2\qquad a=0,1,\dots,r-1,  \qquad m=0,1,\dots.
\end{equation}

Motivated by the $r=3$ solution (\ref{D31}) and (\ref{D33}) we extend  
(\ref{con1}-\ref{con2}) for $a=1,\dots,r-2,\ m=0,1,\dots $ by
\begin{equation}
\bar X^{(a)}_m(x)=y^{(a)}_m\,
\frac
{\big[x^2+\big(p^{(a)}_{m+1}\big)^2\big]
\big[x^2+\big(p^{(a)}_{m-1}\big)^2\big]}
{\big[x^2+\big(p^{(a+1)}_m\big)^2\big] \big[x^2+\big(p^{(a-1)}_m\big)^2\big]}
\label{con3}
\end{equation}
and
\begin{equation}
1+\bar X^{(a)}_m(x)=\big(1+y^{(a)}_m\big)\,
\frac
{\big[x^2+\big(p^{(a)}_m+1\big)^2\big]
\big[x^2+\big(p^{(a)}_m-1\big)^2\big]}
{\big[x^2+\big(p^{(a+1)}_m\big)^2\big] \big[x^2+\big(p^{(a-1)}_m\big)^2\big]}.
\label{con4}
\end{equation}

We are now going to show that it is possible to chose the parameters
$p^{(a)}_m$ so that (\ref{con1}), (\ref{con3}) satisfy the Y-system
(\ref{barX1}). This, together with the consistency between  (\ref{con3}) 
and (\ref{con4}), requires
\begin{equation}
\left.\begin{aligned}
&\xi^{(r-1)}_m=0\\
&\xi^{(r-2)}_m=1+y^{(r-1)}_m\\
&\xi^{(0)}_m=(m+r-1)^2
\end{aligned}\quad\right\}
\qquad m=0,1,\dots,
\label{Xi1}
\end{equation}
\begin{equation}
\xi^{(a)}_0=(r-1-a)^2\qquad a=0,1,\dots,r-1
\label{Xi2}
\end{equation}
and
\begin{eqnarray}
2\big(1+y^{(a)}_m\big)\big(\xi^{(a)}_m+1\big)&=&
y^{(a)}_m\big(\xi^{(a)}_{m+1}+\xi^{(a)}_{m-1}\big)+
\xi^{(a+1)}_m+\xi^{(a-1)}_m\label{x1},\\
\big(1+y^{(a)}_m\big)\big(\xi^{(a)}_m-1\big)^2&=&
y^{(a)}_m\,\xi^{(a)}_{m+1}\,\xi^{(a)}_{m-1}+
\xi^{(a+1)}_m\,\xi^{(a-1)}_m\label{x2}
\end{eqnarray}
for $a=1,\dots,r-2$,\ \ $m=1,2,\dots$.
Using the \lq\lq boundary conditions" (\ref{Xi1}-\ref{Xi2}), (\ref{x1})
alone is already sufficient to calculate all other $\xi^{(a)}_m$. We have
an overdetermined problem and the question is if there is a consistent 
solution (which is unique, if exists).

In addition we have to require that L\"uscher's constraints (\ref{luscherX}) 
are also satisfied. This gives
\begin{eqnarray}
&&\prod_{a=1}^{r-1}\,\big(1+y^{(a)}_1\big)=2r,\label{l1}\\
&&\xi^{(1)}_1=\frac{(r-2)(r^2-1)}{r}.\label{l2}
\end{eqnarray}
Using the formulae of Appendix C we can see that (\ref{l1}) is indeed
satisfied since
\begin{equation}
\prod_{a=1}^{r-1}\,\big(1+y^{(a)}_1\big)=Q^{(1)}_1=2r.
\end{equation}
With the help of $\xi^{(a)}_m$ and the Q-system elements $Q^{(a)}_m$
introduced in Appendix C we now define new variables
\begin{alignat}{3}
&\eta^{(a)}_m=Q^{(a)}_m\,\xi^{(a)}_m &\qquad a&=0,1,\dots,r-1 
&\qquad m&=0,1,\dots,\\
&\hat Q^{(a)}_m=Q^{(a)}_m &\qquad a&=0,1,\dots,r-2 
&\qquad m&=0,1,\dots,\\
&\hat Q^{(r-1)}_m=\big(Q^{(r-1)}_m\big)^2\,
&\qquad m&=0,1,\dots.
\end{alignat}
In terms of these new variables (\ref{x1}) and (\ref{x2}) together
with the Q-system equations (\ref{Qsystem}) can be written as homogeneous
quadratic equations for $a=1,\dots,r-2$,\ \ $m=1,2,\dots$ as follows.
\begin{equation}
\big(\hat Q^{(a)}_m\big)^2=
\hat Q^{(a)}_{m-1}\,\hat Q^{(a)}_{m+1}+
\hat Q^{(a-1)}_m\,\hat Q^{(a+1)}_m,\qquad\qquad\qquad\qquad\quad\ \, 
\label{hatq}
\end{equation}
\begin{equation}
2\hat Q^{(a)}_m\,\big(\eta^{(a)}_m+\hat Q^{(a)}_m\big)=
\hat Q^{(a)}_{m+1}\,\eta^{(a)}_{m-1}+
\hat Q^{(a)}_{m-1}\,\eta^{(a)}_{m+1}+
\hat Q^{(a+1)}_m\,\eta^{(a-1)}_m+
\hat Q^{(a-1)}_m\,\eta^{(a+1)}_m,
\label{hatx1}
\end{equation}
\begin{equation}
\big(\eta^{(a)}_m-\hat Q^{(a)}_m\big)^2=
\eta^{(a)}_{m-1}\,\eta^{(a)}_{m+1}+
\eta^{(a-1)}_m\,\eta^{(a+1)}_m.\qquad\qquad\qquad\qquad\qquad\ \  
\label{hatx2}
\end{equation}
Further simplification is obtained if we write all equations in terms
of the variables $k^{(a)}_m$ and $\tilde k^{(a)}_m$, where
\begin{equation}
\left.
\begin{aligned}
\hat Q^{(a)}_m&=G^{(a)}_m\,k^{(a)}_m,\\
\eta^{(a)}_m&=G^{(a)}_m\,\tilde k^{(a)}_m.
\end{aligned}\quad\right\}\qquad
\begin{aligned}
a&=0,1,\dots,r-1\\ m&=0,1,\dots
\end{aligned}
\end{equation}
Here 
\begin{equation}
G^{(0)}_m=1,\qquad
G^{(a)}_m=\prod_{b=1}^a\,S^{(b)}_m\quad a=1,\dots,r-1
\end{equation}
and finally
\begin{equation}
S^{(b)}_m=\prod_{j=b}^{g-b}\left(\frac{m+j}{j}\right),\qquad
b=1,\dots,r-1.
\end{equation}
From the solution of the constant Q-system we can read off that
\begin{equation}
k^{(0)}_m=1,\quad k^{(1)}_m=\frac{m+r-1}{r-1}\qquad m=0,1,\dots
\end{equation}
and we can prove that if the variables $k^{(a)}_m$ and $\tilde k^{(a)}_m$
are identified with the ones discussed in Appendix D then the quadratic
equations (\ref{hatq}-\ref{hatx2}) are satisfied. (\ref{hatq}),
(\ref{hatx1}) and (\ref{hatx2}) follow from (\ref{QQ1}), (\ref{QQ3}) and
(\ref{QQ2}) respectively. We can show by using the results in Appendix D
that all the remaining requirements (\ref{Xi1}), (\ref{Xi2}) and
(\ref{l2}) are also satisfied.

To summarize, we have shown that the solution of the infinite volume
Y-system equations for the 1-particle case is indeed of the form 
(\ref{con1}), (\ref{con2}), (\ref{con3}) and (\ref{con4}) and calculated
the constant parameters of the solution explicitly.

\setcounter{equation}{0}
\section{1-particle TBA integral equations}
Having determined the infinite volume solution of the 1-particle Y-system
equations we now turn to the problem of transforming these equations into
TBA integral equations.

We start by writing down the integral equations for $X^{(a)}_m(x)$ 
associated to (\ref{X1}). In this case we can use the special case 
of the TBA lemma and we only need to know the positions of the 
poles and zeroes (if any) of the functions
$X^{(a)}_m(x)$ (for $a=1,\dots,r-1$,\ \ $m=1,2,\dots$) in the complex 
$x$ strip with $\vert{\rm Im}\, x\vert\leq1$. From the explicit solution in
the previous section and noting the inequality (\ref{ineq1}) we conclude
that
\begin{eqnarray}
X^{(r-1)}_m(x)\ {\rm has:}&&{\rm double\ zero\ at\ }x=0\quad m=1,2,\dots,
\label{roots1}\\
X^{(r-2)}_m(x)\ {\rm has:}&&{\rm double\ pole\ at\ }x=0\quad m=1,2,\dots,
\label{roots2}\\
X^{(r-2)}_1(x)\ {\rm has:}&&{\rm zeroes\ at\ }x=\pm i.\label{roots3}
\end{eqnarray}
More precisely, we have established (\ref{roots1}-\ref{roots3}) for the
infinite volume solution only. However, since in all other examples
studied previously we found that the distribution of poles/zeroes does 
not depend on the volume we shall assume in the following that 
(\ref{roots1}-\ref{roots3}) hold for all values of the volume $\ell$ and
that no other poles/zeroes occur in the strip $\vert{\rm Im}\, x\vert\leq1$.

This structure can be taken into account by writing
\begin{equation}
X_0(x)={\rm e}^{-\ell\cosh\frac{\pi x}{g}}\,{\rm e}^{\beta_0(x)}
\end{equation}
and
\begin{equation}
\left.
\begin{aligned}
X^{(r-1)}_m(x)&=\tau^2(x)\,{\rm e}^{\beta^{(r-1)}_m(x)}\\
X^{(r-2)}_m(x)&=
\frac{1}{\tau^2(x)}\,{\rm e}^{\beta^{(r-2)}_m(x)}\\
({\rm for\ }r\geq4)\quad
X^{(a)}_m(x)&={\rm e}^{\beta^{(a)}_m(x)}\quad a=1,\dots,r-3\\
\end{aligned}\quad\right\}\qquad m=1,2,\dots,
\end{equation}
where
\begin{equation}
\tau(\theta)=\tanh\left(\frac{\pi\theta}{4}\right).
\end{equation}
The right hand side of (\ref{X1}) is regular for real $x$ and has a double
zero at $x=0$ only for $a=r-2$, $m=1$. Let us introduce
\begin{equation}
\left.
\begin{aligned}
\ell^{(r-1)}_m(x)&=\ln\big\{1+
\tau^2(x)\,{\rm e}^{\beta^{(r-1)}_m(x)}\big\}\\
\ell^{(r-2)}_m(x)&=\ln\big\{
\tau^2(x)+{\rm e}^{\beta^{(r-2)}_m(x)}\big\}\\
({\rm for\ }r\geq4)\quad
\ell^{(a)}_m(x)&=\ln\big\{1+{\rm e}^{\beta^{(a)}_m(x)}\big\}
\quad a=1,\dots,r-3\\
\end{aligned}\quad\right\}\qquad m=1,2,\dots
\end{equation}
and
\begin{equation}
\ell^{(a)}_0(x)=\delta^a_1\,L_0(x)-2\delta^a_{r-2}\,\ln\big(
1+\cosh\frac{\pi x}{2}\big), \qquad a=1,\dots,r-1,
\label{ell0}
\end{equation}
where
\begin{equation}
L_0(x)=\ln\big\{1+{\rm e}^{-\ell\cosh\frac{\pi x}{g}}\,{\rm e}^{\beta_0(x)}
\big\}.
\end{equation}
In the TBA equations we will use the combinations
\begin{equation}
d^{(a)}_m(x)=\ell^{(a)}_{m+1}(x)+\ell^{(a)}_{m-1}(x)+\sum_{b=1}^{r-1}
\,\tilde I_{ab}\,\big[\beta^{(b)}_m(x)-\ell^{(b)}_m(x)\Big]
\end{equation}
defined for $a=1,\dots,r-1$ and $m=1,2,\dots$, where
\begin{equation}
\tilde I_{r-2\,r-1}=2,\qquad \tilde I_{r-1\,r-2}=1,
\end{equation}
otherwise $\tilde I_{ab}$ is the same as $I_{ab}$.

Putting everything together, we can write down the 1-particle TBA equations
\begin{equation}
\beta^{(a)}_m(x)=\delta_{m1}\,\delta^a_{r-2}\,\ln\cosh\frac{\pi x}{2}+
\frac{1}{4}\,\int_{-\infty}^\infty\,{\rm d}u\,\frac{d^{(a)}_m(u)}
{\cosh\frac{\pi(x-u)}{2}}
\label{TBA1}
\end{equation}
for $a=1,\dots,r-1$ and $m=1,2,\dots$.
Note that terms proportional to $\delta^a_{r-2}$ here and in
(\ref{ell0}) take care of the zeroes at $x=\pm i$ for $X^{(r-2)}_1(x)$
and simultaneously for the factor $\tau^2(x)$ on the right hand side of
the $a=r-2$ $m=1$ Y-system equation.

For the TBA integral equation associated to $X_0(x)$ we need the TBA
lemma in its general form. The following observations are necessary
to be able to transform (\ref{X2}) into an integral equation.
\begin{equation}
X_0(x) {\rm \ \ has\ zeroes\ at\ } x=\pm ip^{(1)}_1\qquad
({\rm in\ the\ strip\ }\vert{\rm Im}\,x\vert\leq r-1),
\label{obs1}
\end{equation}
\begin{equation}
\left.
\begin{aligned}
1+X^{(a)}_1(x)\ \ {\rm has\ zeroes\ at\ } x&=\pm i\big(p^{(a)}_1-1\big)\\
{\rm poles\ at\ } x&=\pm ip^{(a+1)}_1
\end{aligned}\right\}
\begin{aligned}
&\qquad{\rm in\ the\ strip}\\
&\ \ \vert{\rm Im}\, x\vert\leq r-1-a\\
&\ \ \ (a=1,\dots,r-2),
\end{aligned}
\label{obs2}
\end{equation}
\begin{equation}
1+X^{(r-1)}_1(x):\ \ {\rm regular\ for\ real\ } x.
\label{obs3}
\end{equation}
(\ref{obs1}) follows from (\ref{Luschery}), (\ref{htheta}) and
(\ref{l2}). (\ref{obs2}) and (\ref{obs3}) follow from the explicit
solution (\ref{con4}) and (\ref{con2}), respectively, taking into account
the inequalities (\ref{ineq1}) and (\ref{ineq2}). As before, we can prove
this analytic structure for the infinite volume case only. We assume the
same analytic structure holds (at least qualitatively) for all values of the
volume parameter $\ell$.

In the notation of Appendix A we have
\begin{equation}
\hat r_a=a-1+p^{(a)}_1,\qquad a=1,\dots,r-2
\end{equation}
for the complex zeroes and similarly
\begin{equation}
\hat p_a=a+p^{(a+1)}_1,\qquad a=1,\dots,r-3
\end{equation}
for the complex poles\footnote{$p^{(r-1)}_1=0$ and consequently
$1+X^{(r-2)}_1(x)$ has a double pole at $x=0$.}.
The product of the relevant $B$-factors is
\begin{equation}
B(x)=\prod_{a=1}^{r-2}\,
\tau\left(\frac{x+i(a-1+p^{(a)}_1)}{r-1}\right)\,
\tau\left(\frac{x-i(a-1+p^{(a)}_1)}{r-1}\right)
\end{equation}
and similarly for the $\tilde B$-factors:
\begin{equation}
\tilde B(x)=\prod_{a=1}^{r-3}\,
\tau\left(\frac{x+i(a+p^{(a+1)}_1)}{r-1}\right)\,
\tau\left(\frac{x-i(a+p^{(a+1)}_1)}{r-1}\right).
\end{equation}
The prefactor in (\ref{Lemma}) becomes
\begin{equation}
\frac{\tilde B(x)}{B(x)}=\frac{1}{
\tau\Big(\frac{x+ip^{(1)}_1}{r-1}\Big)\,
\tau\Big(\frac{x-ip^{(1)}_1}{r-1}\Big).}
\label{pref}
\end{equation}
Remember this prefactor is needed for a solution of the equation
that is analytic and non-vanishing in the strip 
$\vert{\rm Im}\,x\vert\leq r-1$. The poles explicitly present in (\ref{pref})
are canceled by the corresponding zeroes coming from the analytic
continuation of the integral in the exponent. But according to (\ref{obs1}) 
we are looking for a solution that has zeroes at exactly the same positions! 
Thus in the solution for $X_0(x)$ this prefactor is absent and the correct
1-particle TBA integral equation is simply
\begin{equation}
\beta_0(x)=\sum_{a=1}^{r-1}\,\int_{-\infty}^\infty\,
{\rm d}u\,{\cal R}(r-1,r-1-a;x-u)\,\ell^{(a)}_1(u).
\label{TBA2}
\end{equation}

To summarize, the 1-particle TBA integral equations for the O$(2r)$ non-linear
$\sigma$-model are (\ref{TBA1}) and (\ref{TBA2}). The finite volume
1-particle energy in this model is given by
\begin{equation}
E_1(R)= M-\frac{M}{2g}\int_{-\infty}^\infty\,{\rm d}x\,
\cosh\frac{\pi x}{g}\,L_0(x). 
\label{final}
\end{equation}

\setcounter{equation}{0}
\section{Numerical results}

\begin{table}
\begin{center}
\begin{tabular}[t]{c||c|c|c}
$\ell$ & TBA & PT & L\"uscher \\
\hline
\hline
0.001 & 0.401699(7) & 0.40170\\
\hline
0.003 & 0.454323(7) & 0.45432\\
\hline
0.01 & 0.530922(3) & 0.53091\\
\hline
0.03 & 0.628234(3) & 0.62821\\
\hline
0.1 & 0.7881710(6) & 0.78810 & 0.3159\\
\hline
0.3 & 1.0318308(5) & 1.03165 & 0.6409 \\
\hline
1 & 1.58529672(4) & & 1.4128 \\
\hline
2 & 2.3259656(5) & & 2.2838 \\
\hline
\end{tabular}
\end{center}
\caption{{\footnotesize Results for the finite volume mass gap $z$
in the $O(6)$ non-linear $\sigma$-model.}}
\end{table}

In this paper we proposed TBA integral equations to calculate the finite
volume 1-particle energy in the O$(n)$ non-linear $\sigma$-models for
even $n$. This, together with the corresponding equations for the ground 
state \cite{Fen}, enables us to compute the mass gap of the model in
finite volume exactly. Because we do not know how to derive these equations 
from first principles, the proposed equations remain conjectures. For this
reason, it is extremely important to compare the resulting mass gap with
the results of other approaches to the problem.

We have computed the finite volume 1-particle energies together with
the corresponding ground state values for a number of $\ell$ values
numerically. The results for the dimensionless mass gap
\begin{equation}
z=R\big[E_1(R)-E_0(R)\big]
\end{equation}
as function of the dimensionless size of the system, $\ell$, are given
in Table 1 and 2 for the O$(6)$ and O$(8)$ models respectively. 
The corresponding values calculated from L\"uscher's asymptotic formula
(\ref{Luscher}) and three-loop perturbation theory \cite{Shin} are also
given in the tables.

The numerical computation proceeds as usual by iteration (after the 
discretization of the numerical integrals), starting from the
$\ell\to\infty$ solution.
An extra complication for the numerical calculation of the mass gap 
from our TBA equations is caused by the presence of infinitely many
unknowns. We have used here the same method as was applied to the
case of the O$(4)$ model, which is described in \cite{BH2} in some detail.
This is not repeated here. Also the PT formulae necessary to compute the
three-loop perturbative mass gap can be found in \cite{BH2}.

Here we just mention that we checked that to achieve the precision we quote 
it was sufficient to replace the original equations in the numerical 
computation by those of a cutoff problem, which is obtained 
from the TBA equations by assuming that all functions $\beta_0(x)$, 
$\beta^{(a)}_m(x)$ are equal to their infinite volume limit for $x>100$ 
or $m>140$. For sufficient convergence, we need about 10-12 thousand
iterations in these computations, which require about 7-8 hours of CPU
time on a fast PC for the O$(6)$ problem, and almost twice as much for
O$(8)$.

From Table 1 and 2 we see that our numerical results agree with L\"uscher's
asymptotic formula for large volumes and agree with the results of PT
for small volumes. Since our considerations started with L\"uscher's formula,
the large $\ell$ asymptotics is built in, but the very nice four-digit 
agreement between TBA and PT for small $\ell$ is non-trivial. This agreement
makes us confident that our TBA equations are indeed the exact answer to the
question of the finite volume mass gap problem in non-linear $\sigma$-models.


\begin{table}
\begin{center}
\begin{tabular}[t]{c||c|c|c}
$\ell$ & TBA & PT & L\"uscher \\
\hline
\hline
0.001 & 0.38677(2) & 0.3868\\
\hline
0.003 & 0.43870(2) & 0.4387\\
\hline
0.01 & 0.51468(2) & 0.5147\\
\hline
0.03 & 0.61179(2) & 0.6118\\
\hline
0.1 & 0.772630(1) & 0.7727 & 0.2486\\
\hline
0.3 & 1.019830(1) & 1.0200 & 0.5536 \\
\hline
1 & 1.584599(1) & &1.3465 \\
\hline
2 & 2.333139(1) & & 2.2656 \\
\hline
\end{tabular}
\end{center}
\caption{{\footnotesize Results for the finite volume mass gap $z$
in the $O(8)$ non-linear $\sigma$-model.}}
\end{table}





\vspace{1cm}
{\tt Acknowledgements}

\noindent 
This investigation was supported in part by the 
Hungarian National Science Fund OTKA (under T043159 and T049495) and 
by INFN Grant TO12.
A. H. acknowledges the financial support provided through the
European Community's Human Potential Programme under contract
HPRN-CT-2002-00325, \lq EUCLID'.
J. B. wishes to thank the Alexander von Humboldt Foundation for
financial support.

\vspace{2cm}

\setcounter{equation}{0}
\renewcommand{\theequation}{\Alph{section}.\arabic{equation}}
\appendix
\section{TBA Lemma}

\noindent
In this appendix we formulate a simple lemma \cite{Kun} 
that we use to transform TBA Y-systems into integral equations and vice versa.

Let the parameters $\alpha$ and $\beta$ satisfy $0<\beta<\alpha$ and let 
$f(\theta)$ be a (real) meromorphic function in the strip 
$\vert{\rm Im}\,\theta\vert\leq\beta$ with no poles or zeroes 
at the boundaries. Let us denote by
\begin{equation}
\begin{split}
\left\{r_i\right\}\qquad &{\rm the\ set\ of\ real\ zeroes,}\\
\left\{p_j\right\}\qquad &{\rm the\ set\ of\ real\ poles,}\\
\left\{r_\mu\pm ir_\mu^\prime\right\}\quad 
&{\rm the\ set\ of\ complex\ zeroes\ in\ the\ strip}\ \ \ 
(0<r_\mu^\prime<\beta),\\ 
\left\{p_\nu\pm ip_\nu^\prime\right\}\quad 
&{\rm the\ set\ of\ complex\ poles\ in\ the\ strip}\ \ \ 
(0<p_\nu^\prime<\beta). 
\end{split}
\end{equation}
Let us also introduce
\begin{equation}
\hat r_\mu=\alpha-\beta+r_\mu^\prime \qquad\qquad{\rm and}\qquad\qquad
\hat p_\nu=\alpha-\beta+p_\nu^\prime.
\end{equation}
Now we define
\begin{equation}
A(\theta)=\prod_i\,\tau\left(\frac{\theta-r_i}{\beta}\right)\,\qquad\qquad
\qquad
\tilde A(\theta)=\prod_j\,\tau\left(\frac{\theta-p_j}{\beta}\right)
\end{equation}
and
\begin{equation}
\begin{split}
B(\theta)&=\prod_\mu\,
\tau\left(\frac{\theta-r_\mu+i\hat r_\mu}{\alpha}\right)\,
\tau\left(\frac{\theta-r_\mu-i\hat r_\mu}{\alpha}\right),\\
\tilde B(\theta)&=\prod_\nu\,
\tau\left(\frac{\theta-p_\nu+i\hat p_\nu}{\alpha}\right)\,
\tau\left(\frac{\theta-p_\nu-i\hat p_\nu}{\alpha}\right),
\end{split}
\end{equation}
where $\tau(\theta)=\tanh(\pi\theta/4)$.
\begin{equation}
\end{equation}
The function
\begin{equation}
\hat f(\theta)=\frac{\omega\tilde A(\theta)}{A(\theta)}\,f(\theta),
\end{equation}
where $\omega$ is the sign of $f(\infty)$, is positive along the real axis
and we can define
\begin{equation}
F(\xi)=\frac{\tilde B(\xi)}{B(\xi)}\,\exp\left\{
\int_{-\infty}^\infty {\rm d}u\,{\cal R}(\alpha,\beta;\xi-u)\,\ln\hat f(u)
\right\},
\label{Lemma}
\end{equation}
with the kernel function 
\begin{equation}
{\cal R}(\alpha,\beta;\theta)=\frac{1}{\alpha}\,\,
\frac{
\cosh\frac{\theta\pi}{2\alpha}\cdot
\cos\frac{\beta\pi}{2\alpha}}{
\cosh\frac{\theta\pi}{\alpha}+
\cos\frac{\beta\pi}{\alpha}}.
\label{kernel}
\end{equation}

Now the TBA lemma, which can be proved by explicit integration, states that
the function $F(\theta)$ can be analytically continued into the strip
$\vert{\rm Im}\,\theta\vert\leq\alpha$, it is real analytic and non-vanishing (ANZ) 
there and satisfies the functional equation
\begin{equation}
F(\theta+i\alpha)\,F(\theta-i\alpha)=
f(\theta+i\beta)\,f(\theta-i\beta).
\end{equation}

A special (limiting) case is $\beta=0$. In this case the lemma states that
if $\phi(\theta)$ is real analytic in the neighbourhood of the real axis then
the function defined by
\begin{equation}
g(\xi)=\frac{1}{4}\,\int_{-\infty}^\infty {\rm d}u\,\frac{\phi(u)}
{\cosh\frac{\pi}{2}(\xi-u)}=\frac{1}{2}\,\int_{-\infty}^\infty {\rm d}u\,
{\cal R}(1,0;\xi-u)\phi(u)
\end{equation}
is real analytic in the strip $\vert{\rm Im}\,\theta\vert\leq1$ and satisfies
\begin{equation}
g(\theta+i)+g(\theta-i)=\phi(\theta).
\end{equation}

\setcounter{equation}{0}
\section{O$(n)$ S-matrix}
In this appendix we study the properties of the function
\begin{equation}
q(\theta)=\sum_{b=1}^n\,S^{ab}_{ab}(\theta),
\label{Luscherq}
\end{equation}
which appears in L\"uscher's asymptotic formula. 
The matrix elements of the O$(n)$
S-matrix are given by \cite{Zamo}
\begin{equation}
S^{ab}_{cd}(\theta)=\sigma_1(\theta)\,\delta^{ab}\delta^{cd}+
\sigma_2(\theta)\,\delta^{ac}\delta^{bd}+
\sigma_3(\theta)\,\delta^{ad}\delta^{bc},
\end{equation}
where
\begin{equation}
\begin{split}
\sigma_1(\theta)&=-\frac{2\pi i\theta}{i\pi-\theta}\,\,
\frac{S^{(2)}(\theta)}{(n-2)\theta-2\pi i},\\
\sigma_2(\theta)&=(n-2)\theta\,\,
\frac{S^{(2)}(\theta)}{(n-2)\theta-2\pi i},\\
\sigma_3(\theta)&=-2\pi i\,\,
\frac{S^{(2)}(\theta)}{(n-2)\theta-2\pi i}
\end{split}
\end{equation}
and
\begin{equation}
S^{(2)}(\theta)=-\exp\left\{2i\int_0^\infty \frac{{\rm d}\omega}{\omega}\,
\sin(\theta\omega)\,\,\frac{{\rm e}^{-\pi\omega}+
{\rm e}^{-2\pi\frac{\omega}{n-2}}}{1+{\rm e}^{-\pi\omega}}\right\}.
\label{S2}
\end{equation}
From this we get for the combination (\ref{Luscherq})
\begin{equation}
q(\theta)=\frac{S^{(2)}(\theta)\,h\left(\theta-\frac{i\pi}{2}\right)}
{(\theta-i\pi)\left[(n-2)\theta-2\pi i\right]},
\label{Luscherq2}
\end{equation}
where
\begin{equation}
h(\theta)=n(n-2)\theta^2+\frac{\pi^2}{4}\,(n+2)(n-4).
\label{htheta}
\end{equation}

Splitting the Fourier kernel in (\ref{S2}) into two parts we can write
\begin{equation}
S^{(2)}(\theta)=-K(\theta)\,M(\theta),
\end{equation}
where
\begin{equation}
K(\theta)=\exp\left\{2i\int_0^\infty \frac{{\rm d}\omega}{\omega}\,
\sin(\theta\omega)\,\,\left[{\rm e}^{-\pi\omega}+
{\rm e}^{-2\pi\frac{\omega}{n-2}}\right]\right\}
\label{K}
\end{equation}
and
\begin{equation}
M(\theta)=\exp\left\{-i\int_0^\infty \frac{{\rm d}\omega}{\omega}\,
\sin(\theta\omega)\,\,\left[{\rm e}^{-\frac{3\pi\omega}{2}}+
{\rm e}^{-\frac{\pi\omega}{2}-2\pi\frac{\omega}{n-2}}\right]
\,\frac{1}{\cosh\frac{\pi\omega}{2}}\right\}.
\label{M}
\end{equation}
(\ref{K}) can be explicitly evaluated:
\begin{equation}
K(\theta)=\frac{\theta-i\pi}{\theta+i\pi}\,\,
\frac{(n-2)\theta-2\pi i}{(n-2)\theta+2\pi i}.
\end{equation}
Using this in (\ref{Luscherq2}) we get
\begin{equation}
y(\theta)=q\left(\theta+\frac{i\pi}{2}\right)=
-\frac{M\left(\theta+\frac{i\pi}{2}\right)\,h(\theta)}
{\left(\theta+\frac{3i\pi}{2}\right)\,\left[(n-2)\theta+
\frac{i\pi}{2}(n+2)\right]}.
\label{Luschery}
\end{equation}
$M(\theta)$ cannot be explicitly evaluated in terms of elementary
functions but for our purposes it is sufficient to notice that it is
an analytic and non-vanishing (ANZ) function in the strip
$0\leq{\rm Im}\,\theta\leq\pi$ and satisfies
\begin{equation}
M(\theta)M(\theta+i\pi)=
\frac{\theta+2\pi i}{\theta-i\pi}\,\,
\frac{(n-2)\theta+i\pi n}{(n-2)\theta-2\pi i},
\end{equation}
which can be shown using the Fourier representation (\ref{M}). It then follows
that the combination relevant for the TBA Y-system is a rational function and
is given by
\begin{equation}
y\left(\theta+\frac{i\pi}{2}\right)y\left(\theta-\frac{i\pi}{2}\right)=
\frac{h\left(\theta+\frac{i\pi}{2}\right)h\left(\theta-\frac{i\pi}{2}\right)}
{(\theta^2+\pi^2)\left[(n-2)^2\theta^2+4\pi^2\right]}.
\end{equation}

Finally we will compute the function (\ref{Luschery}) along the real axis. 
We get
\begin{equation}
y(\xi)=\frac{h(\xi)M_0(\xi)}
{\sqrt{\xi^2+\frac{9\pi^2}{4}}\,\sqrt{(n-2)^2\xi^2+
\frac{\pi^2}{4}(n+2)^2}},
\label{Luscheryreal}
\end{equation}
where
\begin{equation}
M_0(\xi)=\exp\left\{\int_0^\infty \frac{{\rm d}\omega}{\omega}\,
\cos(\xi\omega)\,\,\tanh\frac{\pi\omega}{2}
\left[{\rm e}^{-\frac{3\pi\omega}{2}}+
{\rm e}^{-\frac{\pi\omega(n+2)}{2(n-2)}}\right]\right\}.
\label{L0}
\end{equation}

\setcounter{equation}{0}
\section{Constant Y-system and Q-system}
In this appendix we summarize the results for the standard $D_r$ Q-system
\cite{Kun2}. We first study the generic case $r\geq3$ and define for finite 
$k\geq2$ the variables $Q^{(a)}_m$
where the upper index $a=1,\dots,r$ labels the nodes of the $D_r$ 
Dynkin-diagram and the lower index takes values $m=-1,0,\dots,k,k+1$. 
They satisfy the Q-system 
equations  
\begin{equation}
\left(Q^{(a)}_m\right)^2=Q^{(a)}_{m+1}Q^{(a)}_{m-1}+
\prod_{b=1}^r\left(Q^{(b)}_m\right)^{I_{ab}},
\label{Qsystem}
\end{equation}
where $I_{ab}$ is the incidence matrix of the $D_r$ Dynkin 
diagram\footnote{i.e. $I_{ab}=1$ if the nodes $a,b$ are connected and vanishes
otherwise} for $a=1,\dots,r$ and $m=1,\dots,k-1$. By definition
\begin{equation}
Q^{(a)}_0=Q^{(a)}_k=1,\qquad\qquad
Q^{(a)}_{-1}=Q^{(a)}_{k+1}=0,\qquad\qquad a=1,\dots,r.
\end{equation}
To simplify some formulae we will also use the definition $Q^{(0)}_m=1$ 
($m=0,1,\dots,k$).
Due to the reflection symmetry of the $D_r$ Dynkin-diagram we have
$Q^{(r-1)}_m=Q^{(r)}_m$.

From the solution of the Q-system we can construct 
\begin{equation}
y^{(a)}_m=\frac{Q^{(a)}_{m+1}Q^{(a)}_{m-1}}
{\prod_{b=1}^r\left(Q^{(b)}_m\right)^{I_{ab}}},\qquad
1+y^{(a)}_m=\frac{\left(Q^{(a)}_m\right)^2}
{\prod_{b=1}^r\left(Q^{(b)}_m\right)^{I_{ab}}}
\end{equation}
for $a=1,\dots,r$ and $m=0,1,\dots,k$
and it is easy to show that they satisfy the constant version of the standard
$D_r$ Y-system:
\begin{equation}
\left(y^{(a)}_m\right)^2=(1+y^{(a)}_{m+1})(1+y^{(a)}_{m-1})\,
\prod_{b=1}^r\left(\frac{y^{(b)}_m}{1+y^{(b)}_m}\right)^{I_{ab}}
\label{ysystem}
\end{equation}
for $a=1,\dots,r$ and $m=1,\dots,k-1$.

The unique positive solution of the Q-system (\ref{Qsystem}) is 
characterized by \cite{Kun2}
\begin{equation}
Q^{(1)}_m=\frac{p(m+r-1)}{p(r-1)}\,
\prod_{j=1}^{2r-3}\,\frac{p(m+j)}{p(j)},\qquad\quad m=-1,0,\dots,k,k+1,
\label{Q1}
\end{equation}
where 
\begin{equation}
p(n)=\sin\left(\frac{n\pi}{k+g}\right)
\end{equation}
and $g=2(r-1)$ is the Coxeter number of the $D_r$ Lie-algebra. The other
components $Q^{(a)}_m$ for $a=2,\dots,r$ are given by more complicated 
expressions, which can be obtained recursively using the Q-system 
equations and the result ({\ref{Q1}).

For the O$(2r)$ TBA system we need the $k\to\infty$ limit of the Q-system and
the constant Y-system. The solution (\ref{Q1}) in this limit simplifies to
\begin{equation}
Q^{(1)}_m=\frac{m+r-1}{r-1}\,
\prod_{j=1}^{2r-3}\,\frac{m+j}{j},\qquad\quad m=-1,0,\dots.
\label{Q1infty}
\end{equation}

In the first nontrivial case, $r=3$, the solution of the constant Y-system is
\begin{equation}
y^{(1)}_m=\frac{m(m+4)}{4},\qquad\qquad
y^{(2)}_m=\frac{m(m+4)}{3},\qquad m=0,1,\dots.
\end{equation}

The Q-system relevant for the $r=2$ case is
\begin{equation}
(Q_m)^2=Q_{m+1}Q_{m-1}+1
\end{equation}
for $m=0,1,\dots,k$, with boundary conditions $Q_0=Q_k=1$, 
$Q_{-1}=Q_{k+1}=0$. The Y-system is  
\begin{equation}
(y_m)^2=(1+y_{m+1})(1+y_{m-1}),\qquad m=1,\dots,k-1,
\end{equation}
which is solved by
\begin{equation}
y_m=Q_{m+1}Q_{m-1},\qquad\qquad 1+y_m=Q_m^2,\qquad m=0,1,\dots,k.
\end{equation}
In this case the complete solution is
\begin{equation}
Q_m=\frac{p(m+1)}{p(1)},
\end{equation}
which reduces to
\begin{equation}
Q_m=m+1\qquad\qquad{\rm and}\qquad\qquad y_m=m(m+2)
\end{equation}
in the $k\to\infty$ limit.

\setcounter{equation}{0}
\section{Linearization of the Q-system equations}
\noindent
In this appendix we study a system of linear recursion relations that are
useful in finding the solution of the infinite volume TBA Y-system.
As a \lq\lq byproduct" of this construction we find how to linearize the
recursion relations for the infinite Q-system. 

The variables in these recursion relations are
\begin{equation}
k^{(a)}_m,\,\tilde k^{(a)}_m,\qquad a=0,1,\dots,r-1, 
\end{equation}
where $r\geq2$. The \lq\lq initial conditions" are
\begin{equation}
k^{(0)}_m=1,\quad
k^{(1)}_m=\frac{2m+g}{g},\qquad
\tilde k^{(0)}_m=(m+r-1)^2,
\end{equation}
where $g=2(r-1)$, and the rest of the variables are determined recursively
by the relations
\begin{equation}
(2m+g)k^{(a)}_m=(g-a)k^{(a+1)}_m+ak^{(a-1)}_m
\label{D}
\end{equation}
and
\begin{equation}
r(r-1)\tilde k^{(a)}_m=(r-a-1)(m+r-1)\{(r-a)(m+r-1)k^{(a)}_m-
ak^{(a-1)}_m\}.
\label{D1}
\end{equation}
(\ref{D}) is valid for $a=1,\dots,r-2$, while (\ref{D1}) for
$a=1,\dots,r-1$. It is clear from these relations that $k^{(a)}_m$
and $\tilde k^{(a)}_m$ are polynomials in $m$ (of order $a$ and $a+2$
respectively).

Using the definition (\ref{D}) we can prove by induction that the variables
$k^{(a)}_m$ satisfy further linear relations. These are
\begin{equation}
(g-2a)k^{(a)}_m=(g+m)k^{(a)}_{m-1}-mk^{(a)}_{m+1}
\label{*}
\end{equation}
for $a=0,1,\dots,r-1$,
\begin{equation}
(m+g-a)k^{(a)}_m=(g+m)k^{(a)}_{m-1}+ak^{(a-1)}_m
\label{**}
\end{equation}
for $a=1,\dots,r-1$ and
\begin{equation}
(a-g-m)k^{(a)}_m=mk^{(a)}_{m+1}+(a-g)k^{(a+1)}_m
\label{***}
\end{equation}
for $a=0,1,\dots,r-2$. 
Further these relations can be used to show that $\tilde k^{(a)}_m$ can also
be calculated from
\begin{equation}
r(r-1)\tilde k^{(a)}_m=(r-a-1)(m+r-1)\{(r-a-2)(m+r-1)k^{(a)}_m+
(g-a)k^{(a+1)}_m\}.
\label{D2}
\end{equation}
Our variables also satisfy three sets of quadratic relations (for
$a=1,\dots,r-2$). These are
\begin{equation}
(m+a)(m+g-a)(k^{(a)}_m)^2=m(m+g)k^{(a)}_{m+1}k^{(a)}_{m-1}+
a(g-a)k^{(a+1)}_mk^{(a-1)}_m,
\label{QQ1}
\end{equation}
\begin{equation}
(m+a)(m+g-a)(k^{(a)}_m-\tilde k^{(a)}_m)^2=
m(m+g)\tilde k^{(a)}_{m+1}\tilde k^{(a)}_{m-1}+
a(g-a)\tilde k^{(a+1)}_m\tilde k^{(a-1)}_m
\label{QQ2}
\end{equation}
and
\begin{equation}
\begin{split}
2(m+a)(m+g&-a)k^{(a)}_m(k^{(a)}_m+\tilde k^{(a)}_m)=
m(m+g)[k^{(a)}_{m+1}\tilde k^{(a)}_{m-1}+
\tilde k^{(a)}_{m+1}k^{(a)}_{m-1}]\\
&+a(g-a)[k^{(a+1)}_m\tilde k^{(a-1)}_m+
\tilde k^{(a+1)}_mk^{(a-1)}_m].
\end{split}
\label{QQ3}
\end{equation}
We can prove these relations by expressing all variables occurring in them
in terms of $k^{(a)}_{m-1}$ and $k^{(a-1)}_m$ using the linear relations
(\ref{D}),(\ref{D1}),(\ref{*}),(\ref{**}),(\ref{***}) and (\ref{D2}).
As functions of these two variables the quadratic relations (\ref{QQ1}),
(\ref{QQ2}) and (\ref{QQ3}) are identities.

In the main text we need the following special cases.
\begin{eqnarray}
k^{(a)}_0&=&1,\qquad \tilde k^{(a)}_0=(r-a-1)^2\qquad a=0,1,\dots,r-1,
\label{spec1}\\
k^{(r-1)}_m&=&\frac{(m+1)(m+3)\cdots (m+g-1)}{1\cdot3\cdots(g-1)},
\qquad \tilde k^{(r-1)}_m=0,
\label{spec2}\\
\tilde k^{(1)}_m&=&\frac{(r-2)(m+r)(m+r-2)}{r}\,k^{(1)}_m,
\label{spec3}\\
k^{(r-2)}_m&=&\frac{1}{r-1}\{(m+r-1)k^{(r-1)}_m-(m+g)k^{(r-1)}_{m-1}\},
\label{spec4}\\
\tilde k^{(r-2)}_m&=&\frac{m+r-1}{r-1}\,k^{(r-1)}_m.
\label{spec5}
\end{eqnarray}

For $m\geq0$ integer we define the ratio
\begin{equation}
\xi^{(a)}_m=\frac{\tilde k^{(a)}_m}{k^{(a)}_m}.
\end{equation}
From (\ref{spec1}-\ref{spec5}) we now have
\begin{eqnarray}
\xi^{(0)}_m&=&(m+r-1)^2,\ \  \xi^{(1)}_m=\frac{(r-2)(m+r)(m+r-2)}{r},
\ \  \xi^{(r-1)}_m=0,
\label{xi1}\\
\xi^{(a)}_0&=&(r-a-1)^2,\qquad a=0,1,\dots,r-1.
\end{eqnarray}
The variables $\xi^{(a)}_m$ also satisfy the following inequalities.
\begin{equation}
\xi^{(a)}_m\geq1,\qquad\qquad a=0,1,\dots,r-2,\qquad m\geq0,
\label{ineq1}
\end{equation}
where equality is for the case $a=r-2$, $m=0$ only.\\
Also
\begin{equation}
\frac{(r-a-1)^2(m+r-1)(m+r-2)}{r(r-2)}<\xi^{(a)}_m<
\left[\frac{(r-a)(m+r-1)}{r}\right]^2
\end{equation}
for $a=1,\dots,r-2$, $m\geq1$. An important consequence of the above is
\begin{equation}
r-a-1<\sqrt{\xi^{(a)}_1}<r-a,\qquad\quad a=1,\dots,r-2.
\label{ineq2}
\end{equation}

\end{document}